\documentclass{ifacconf}

\usepackage{graphicx}      
\usepackage{epstopdf}
\usepackage{amsmath,amsfonts}
\renewcommand\epsilon{\varepsilon}
\renewcommand\phi{\varphi}
\newcommand{\trn}{^{\rm\scriptscriptstyle T}}
\DeclareMathOperator{\col}{col}
\usepackage{natbib}        
\newtheorem{deff}{Definition}
 \DeclareMathOperator{\diag}{diag}
\begin{document}
\begin{frontmatter}

\title{Desynchronization in Oscillatory Networks Based on Yakubovich Oscillatority} 

\thanks[footnoteinfo]{The analysis of desynchronization problem in oscillatory networks (Theorem~\ref{des}) was performed in Lobachevsky State University of Nizhny Novgorod and supported by Russian Science Foundation (project no. 19-72-10128). \\
The desynchronization conditions for FitzHugh-Nagumo network with diffusive coupling (Theorem~\ref{FHNt}) were obtained in IPME RAS and funded by RFBR according to the research (Project No. 18-38-20037) and by the President of Russian Federation grant for the support of young Russian scientists (Grant No. MK-3621.2019.1).}

\author[First,Second]{Sergei A. Plotnikov} 
\author[First,Third]{Alexander L. Fradkov} 

\address[First]{Institute for Problems of Mechanical Engineering, Russian Academy of Sciences, Bolshoy~Ave 61, Vasilievsky Ostrov, St.~Petersburg, 199178, Russia.}
\address[Second]{Lobachevsky State University of Nizhny Novgorod, Gagarina Pr. 23, Nizhny Novgorod, 603950, Russia (e-mail: waterwalf@gmail.com).}
\address[Third]{Saint Petersburg State University, Universitetskii Pr. 28, St. Petersburg 198504, Russia (e-mail: fradkov@mail.ru).}

\begin{abstract}                
The desynchronization problems in oscillatory networks is considered. A new desynchronization notion is introduced and desynchronization conditions are provided. The desynchronization notion is formulated in terms of Yakubovich oscillatority of the auxiliary synchronization error system. As an example, the network of diffusively coupled FitzHugh-Nagumo systems with undirected graph is considered. The simple inequality guaranteeing network desynchronization is derived. The simulation results confirm the validity of the obtained analytical results.
\end{abstract}


\end{frontmatter}

\section{Introduction}
The synchronization phenomenon in oscillatory networks has been intesnively studied recents years and has a lot of applications in biology, social sciences, physis and engeneering \cite{BLE88,PIK03}. Particularly, it plays important role in neural networks, where it is associated with emergence of pathological rhytmic brain activity inessential tremor \cite{SCH09}, epilepsy \cite{WON86}, and Parkinson's disease \cite{HAM07}.
Normally, neurons fire in an uncorrelated manner, i.e. the neuron network is desynchronized. Motivated by this problem, different works have appeared to study desynchronization \cite{DTC18,ZAN19,TYL19} and to design the control algorithms , such as linear and nonlinear
time delayed feedback \cite{ROS04a,POP05,CHE15}, pulsatile stimulation protocols \cite{TAS01,POP17}, proportional integro-differential feedback \cite{PYR07}, control based on unscented Kalman filter \cite{LU18}, or simple controller based on the inversion of the mean field \cite{HON11,TAM12,PLO19a}. However, the majority of these works mostly rely on numerical analysis.

While there are a lot of different definitions of synchronization (see, e.g. \cite{BLE88,STR00,FRA07,OSI07}), only few attempts were made to formalize desynchronization \cite{FRA12,ZHA13}. However, these definitions are not always appropriate for use. A rigorous mathematical definition of desynchronization was proposed in \cite{FRA11,FRA12}. However, the definition of \cite{FRA11,FRA12} covers only the networks consisting  of coupled phase systems. In \cite{ZHA13} authors define desynchronization as a stability of synchronization error system with different delays. However, this definition also do not cover all possible scenarios of desynchronization, since the delay may not be always constant. In our paper a new definition of desynchronization suitable for more general class of networks is proposed. Our definition is based on the oscillatority notion introduced by Yakubovich \cite{YAK73}.

Within the framework of this approach, the frequency conditions for oscillatority are obtained for the class of so-called Lurie systems, consisting of the nominal linear part and nonlinearity in the feedback circuit \cite{YAK73,YAK75,TOM89,LEO96}. Lyapunov functions based on the quadratic form of the state variables of the linear part of the system are used to study the oscillatority of the system. As the application examples this approach can be used to solve the problems of vibration analysis in control systems \cite{YAK75}, phase systems \cite{LEO96}, and chaotic systems \cite{FRA98}. While in this paper we apply this approach to obtain desynchronization conditions in nonlinear oscillatory systems. Further, we apply this result to the case of the network of one of the simplest neuron models, namely FitzHugh-Nagumo model \cite{FIT61a,NAG62}.

The rest of the paper is organized as follows. In Sec.~\ref{sec:prel} we recall some basic notions and theorems about Yakubovich oscillatority, graph theory, singularly perturbed systems and FitzHugh-Nagumo model. Section~\ref{sec:main} introduces the notion of desynchronization and derives the theorem about its conditions. In Sec.~\ref{sec:example} the network of diffusively coupled FitzHugh-Nagumo systems with undirected graph is considered, desynchronization conditions for this network are derived and simulation results are provided. Finally, we conclude with Sec.~\ref{sec:conclusion}.

{\bf Notation.} Throughout the paper $\mathbb{R}^n$ denotes the $n$ dimensional Euclidean space with vector norm $|\cdot|$; notation $z=\col(x,y)$ means that $z$ is the vector of two components $x,~y$; superscript $\trn$ stands for matrix transposition; notation $P\succeq0$ for matrix $P$ means that $P$ is symmetric and positive semidefinite; $A\otimes B$ means the Kronecker product of matrices $A$ and $B$.

\section{Preliminaries}\label{sec:prel}
\subsection{Yakubovich Oscillatority}
Consider the following nonlinear system of Lurie type:
\begin{equation}\label{f2}
\begin{aligned}
\dot x(t)=& Ax(t)+B \phi\left(y(t)\right), \\
 y(t) =&Rx(t),
 \end{aligned}
\end{equation}
where $x \in \mathbb{R}^n$ is a state vector, $y\in \mathbb{R}^p$ is an output, $A$, $B$, $R$ are the matrices of the appropriate dimensions, $\phi:\mathbb{R}^{p}\to\mathbb{R}^{k}$ is a smooth function. Let us note that any nonlinear system $\dot x=f(x)$ can be written in the form \eqref{f2}. This form, in which the linear part is explicitly singled out, is usual in works dealing with absolute stability. 

 Remind some basic definitions about Yakubovich oscillatority \cite{TOM89,EFI06}.
\begin{deff}
For $\alpha,\beta \in \mathbb{R}$ and $\alpha<\beta$ the function $\psi:\mathbb{R}\to\mathbb{R}$ is $(\alpha,\beta)$-oscillatory while $t\to\infty$ if it is bounded and the following inequalities are fulfilled:
$$
\limsup_{t\to\infty}\psi(t)\ge\beta,\quad\liminf_{t\to\infty}\psi(t)\le\alpha.
$$

The function $\psi:\mathbb{R}\to\mathbb{R}$ is oscillatory in the sense of Yakubovich while $t\to\infty$ if there exist some constants $\alpha$ and $\beta$ such that function $\psi(t)$ is $(\alpha,\beta)$-oscillatory while $t\to\infty$.

The system \eqref{f2} is oscillatory in the sense of Yakubovich by output $\psi=\eta(x)$ if for almost all $x_0\in\mathbb{R}^n$ the system solutions $x(t,x_0)$ are bounded and for almost all initial conditions the following inequality holds:
\begin{equation}\label{f3}
\liminf_{t\to\infty}\psi(x(t,x_0))<\limsup_{t\to\infty}\psi(x(t,x_0)).
\end{equation}
\end{deff}

The property \eqref{f3} should be fulfilled for all solutions because of the possibility of existence of null empty set of initial conditions, for which the corresponding solution of the system \eqref{f2} is not an oscillation.

\begin{thm}\label{osc}
Let all solutions of the system \eqref{f2} be bounded, and all matrices $Q_j=A+  B\partial\phi(Rx_j^*)/\partial y$ have at least one eigenvalue with positive real part and do not have imaginary eigenvalues, where $x_j^*$ is an equilibrium point of the system \eqref{f2} (i.e. it is a solution of the equation $A x^*+ B\phi(Rx^*)=0$). Then the system \eqref{f2} is oscillatory in the sense of Yakubovich. If all the eigenvalues of the matrices $Q_j$ have positive real parts then the set of initial conditions $\Omega$, for which the solutions are not the oscillations by output $y$ while $t\to\infty$, has the form $\Omega=\{x:x=x_j^*\}$.
\end{thm}

\subsection{Graph Theory}
Here we remind some definitions from graph theory.
\begin{deff}
An undirected graph is an ordered pair $G=(V,A)$, where $V$ is non-empty set of vertices or nodes, and $A\subseteq V\times V$ is a set of unordered pairs of distinct nodes, which are called arcs or edges. 
\end{deff}
\begin{deff}
A weighted graph is a graph in which each edge is given a numerical weight.
\end{deff}
\begin{deff}
An adjacency matrix $A=(\gamma_{ij}(G))$ of graph $G$ of $n$ nodes is a $n\times n$ matrix those $ij$th entry is $w\ne0$ if node $i$ connects to node $j$ by the edge with the weight equals $w$, and is $0$ otherwise.
\end{deff}
An adjacency matrix of undirected graph is always symmetric. 
\begin{deff}
The Laplacian matrix of graph $G$ of $n$ nodes is a $n\times n$ matrix which is given by
$$
L(G)=\begin{bmatrix} 
\sum_{j=2}^{n}\gamma_{1j} &-\gamma_{12} &\cdots&-\gamma_{1n}\\
-\gamma_{21} &\sum_{j=1,i\ne j}^{n}\gamma_{2j}&\cdots&-\gamma_{2n} \\
\vdots &\vdots &\ddots&\vdots \\
-\gamma_{n1} &-\gamma_{n2} &\cdots&\sum_{j=1}^{n-1}\gamma_{nj}
\end{bmatrix},
$$
where $\gamma_{ij}$ are the elements of its adjacency matrix.
\end{deff}
If the graph $G$ is undirected, then $L(G)=L(G)\trn$ and $\lambda_1(G)$ (the least eigenvalue of $L(G)$) is equal to $0$, particularly $L(G)\succeq 0$ \cite{OLF07,AGA09}.

\subsection{Singularly Perturbed System}
Consider the singulalry perturbed linear system, which is the system in which the derivatives of some states are multiplied by a small positive scalar $\epsilon$, i.e.
\begin{equation}\label{f1}
\begin{aligned}
\dot x_1(t) &= A_{11}x_1(t)+A_{12}x_2(t), \\
\epsilon \dot x_2(t) &= A_{21}x_1(t)+A_{22}x_2(t),
\end{aligned}
\end{equation}
where $x_1(t)\in \mathbb{R}^{n_1}$, $x_2(t)\in \mathbb{R}^{n_2}$ are the state vectors, $A_{ij}$ ($i=1,2$, $j=1,2$) are the matrices of the appropriate dimensions. For this system the following proposition holds (see e.g. \cite{ KOK86}):
\begin{prop}\label{sing}
Let the following assumptions be fulfilled:
\begin{enumerate}
\item
The ``fast'' matrix $A_{22}$ is Hurwitz.
\item
The ``slow'' matrix $A_{11}-A_{12}A_{22}^{-1}A_{21}$ is Hurwitz.
\end{enumerate}
Then the system \eqref{f1} is robustly asymptotically stable.
\end{prop}

\subsection{FitzHugh-Nagumo Model}\label{sec:model}
Consider one of the simplest models of neuron dynamics which is presented as a form of two-dimensional oscillator with a cubic nonlinearity and called FitzHugh-Nagumo (FHN) model \cite{IZH10,FIT61a,NAG62}. 

\begin{figure}
\includegraphics[width=1\linewidth]{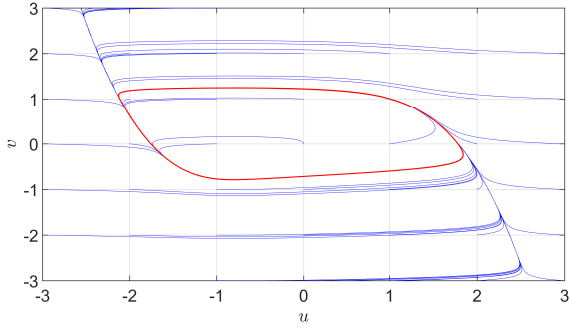}
\caption{Phase portrait of FHN system \eqref{m}. The system trajectories are marked by blue color, while limite cycle is marked by red color. Parameters: $\epsilon=0.1$, $a=0.8$.}
\label{figs}
\end{figure}

\begin{equation}\label{m}
\begin{aligned}
\epsilon \dot u(t)&=u(t)-\frac{u^3(t)}{3}-v(t), \\
\dot v(t)&=u(t)+a,
\end{aligned}
\end{equation}
where $u$ and $v$ are the components of the state vector meaning the membrane potential (activator) and recovery variable (inhibitor) of the neuron, respectively; $\epsilon$ is a time-scale parameter, which separates the fast and slow dynamics and is typically smal, i.e. $0<\epsilon\ll1$; $a$ is a threshold parameter: for $|a|>1$ the system is excitable, while for $|a|<1$ it is oscillatory. This is due to a supercritical Andronov-Hopf bifurcation at $|a|=1$ with a locally stable equilibrium point for $|a|>1$ and a stable limit cycle for $|a|<1$.

Figure~\ref{figs} depicts the phase portrait of FHN model with parameters $\epsilon=0.1$, $a=0.8$. Starting from arbitrary points the system trajectories converge to the limit cycle (marked by red color), i.e. for chosen parameters, FHN system is oscillatory.

\section{Main Result}\label{sec:main}
Consider the network of $N$ diffusively coupled identical nonlinear systems in normal form:
\begin{equation}\label{f4}
\begin{aligned}
\dot y_i=& A_yx_i+ b_y\phi_y(rx_i) + \sum_{j=1}^{N}\gamma_{ij}\left[y_j-y_i\right], \\
\dot z_i=&A_z x_i+ b_z\phi_z\left(rx_i\right),
 \end{aligned}
\end{equation}
where $y_i\in \mathbb{R}^m$, $z_i\in \mathbb{R}^{n-m}$, $x_i=\col(y_i,z_i)\in\mathbb{R}^n$ are the state variables, particularly $z_i$ corresponds to the $i$th node's zero-dynamics \cite{ISI99}. $A_y$, $A_z$, $b_y$, $b_z$, $r$ are the matrices of the appropriate dimensions; the functions $\phi_y:\mathbb{R}^{p}\to \mathbb{R}^{k_y}$, $\phi_z:\mathbb{R}^{p}\to \mathbb{R}^{k_z}$ are smooth; $\gamma_{ij}\ge0$ are coupling coefficients. Denote the state of the network as $x=\col(x_1,\dots,x_N)\in\mathbb{R}^{Nn}$ and suppose that the graph $G$ of considered network is undirected, i.e. its Laplacian matrix $L(G)$ is symmetric.

Introduce a compact notation for the system \eqref{f4}:
$$
\begin{gathered}
A=I_N\otimes \begin{bmatrix}A_y\\A_z \end{bmatrix},~ B=I_N\otimes\begin{bmatrix}b_y\\b_z \end{bmatrix}, ~ R=I_N\otimes r,\\
 E_m=\begin{bmatrix}I_m\\0_{m\times(n-m)} \end{bmatrix},~
H=-[L(G)\otimes E_m][I_N\otimes E_m\trn], \\
\phi(Rx)=\begin{bmatrix}\phi_y(rx_1)\\ \phi_z(rx_1)\\\vdots\\\phi_y(rx_N)\\ \phi_z(rx_N)\end{bmatrix},
\end{gathered}
$$
where $A$ and $H$ is $Nn\times Nn$-matrices, $B$ is $Nn\times Nk$-matrix, $R$ is $Np\times Nn$ matrix, $E_m$ is $n\times m$ matrix, $\phi:\mathbb{R}^{Np}\to\mathbb{R}^{Nk}$, $k=k_y+k_z$. With this notation the network \eqref{f4} can be rewritten in the form:
\begin{equation}\label{f6}
\dot x = (A+H)x+B\phi (Rx).
\end{equation}

Now we transform system \eqref{f4} to new coordinates $x_s=\col(\bar x, e_s)=(U\otimes I_n)x$, using matrix $U\otimes I_n$, where
$$
U=\begin{bmatrix}
\phantom{-}\frac{1}{N}&\phantom{-}\frac{1}{N}&\dots&\phantom{-}\frac{1}{N}&\phantom{-}\frac{1}{N}\\
\frac{N-1}{N}&-\frac{1}{N} &\dots&-\frac{1}{N}&-\frac{1}{N}\\
-\frac{1}{N}&\frac{N-1}{N}&\ddots&-\frac{1}{N}&-\frac{1}{N}\\
\vdots &\vdots&\ddots&\ddots&\vdots\\
-\frac{1}{N}&-\frac{1}{N}&\dots&\frac{N-1}{N}&-\frac{1}{N} \\
\end{bmatrix},
$$
and $I_n$ is an identity $n\times n$-matrix; $\bar x=\col(\bar y,\bar z) \in \mathbb{R}^n$ describes the mean field, while $e_s =\col (e_{y1}, e_{z1},\dots,e_{y(N-1)},$ $e_{z(N-1)})  \in \mathbb{R}^{N(n-1)}$ describes the differences between the $i$th node's state and the mean field, which are the synchronization errors. Since the graph $G$ is undirected and $y_j-y_i=y_j-\bar y+\bar y-y_i=e_{yj}-e_{yi}$, the network in new coordinates is presented as follows:
 \begin{subequations}
 \begin{equation}\label{f5a}
\begin{aligned}
\dot {\bar y}=& A_y\bar x+ \frac{1}{N}b_y\sum\limits_{j=1}^{N}\phi_y(rx_j), \\
\dot {\bar z}=& A_z \bar x+ \frac{1}{N} b_z\sum\limits_{j=1}^{N}\phi_z(rx_j), 
\end{aligned}
\end{equation}
\begin{equation}\label{f5b}
\begin{aligned}
\dot e_{yi}=& A_ye_{yi}+ b_y\left(\phi_y(rx_i)-  \frac{1}{N}\sum\limits_{j=1}^{N}\phi_y(rx_j)\right)\\
&+ \sum_{j=1}^{N}\gamma_{ij}\left[e_{yj}-e_{yi}\right], \\
\dot e_{zi}=& A_z e_{zi}+ b_z\left(\phi_z(rx_i)-  \frac{1}{N}\sum\limits_{j=1}^{N}\phi_y(rx_j)\right).
 \end{aligned}
 \end{equation}
 \end{subequations}

To study synchronization one can analyze the stability of the synchronization error system \eqref{f5b}. The stability of this system means the convergence of the synchronization errors to zero, which, in its turn, means the convergence the i$th$ node's state to the mean field, i.e. network \eqref{f4} synchronization. The reasonable question is how to define a desynchronization in a network. There are some definitions of desynchronization \cite{FRA11,ZHA13}, however, they do not cover all possible scenarios of desynchronization. Here we introduce different definition in terms of synchronization errors:
\begin{deff}
The network \eqref{f4} is called desynchronized by output $\psi=\eta(x)$ if the corresponding synchronization error system \eqref{f5b} is oscillatory in the sense of Yakubovich by output $\psi=\eta(x)$.
\end{deff}
This means that synchronization errors are bounded and oscillate. To find the desynchronization conditions one can apply Theorem~\ref{osc} to synchronization error system \eqref{f5b}. Firstly, one should suppose that all solutions of system \eqref{f5b} are bounded. It is equivalent to the boundedness of the solutions of original system \eqref{f4} solutions. The second step is to find all equilibrium points of the system \eqref{f5b} and calculate the matrices $Q_j$. The eigenvalues of the matrices do not depend on the coordinate transformation. Therefore, one can directly check the Theorem~\ref{osc} conditions for the original network \eqref{f4}.

Thus, the following theorem for network \eqref{f6} desynchronization can be formulated:
\begin{thm}\label{des}
Let all solutions of the system \eqref{f6} be bounded, and all the eigenvalues of matrices $Q_j=(A+H)+B\partial\phi(Rx_j^*)/\partial (Rx)$ have positive real parts, where $x_j^*$ is an equilibrium point of the system \eqref{f6} (i.e. it is a solution of the equation ($(A+H) x^*+ B\phi(Rx^*)=0$). If initial conditions do not belong to the set $\Omega=\{x:x= x_j^*\}$, then the network \eqref{f6} is desynchronized by output $Rx$.
\end{thm}
Note that we assume that the graph is undirected to derive the synchronization error system. However, we do not assume this in the Theorem~\ref{des}.

\section{Example}\label{sec:example}
As an example consider the network of $N$ diffusively coupled FHN systems with undirected graph:
\begin{equation}\label{f7}
\begin{aligned}
\epsilon \dot u_i&=u_i-\frac{u_i^3}{3}-v_i+\sum\limits_{j=1}^{N}\gamma_{ij}\left[u_j-u_i\right], \\
\dot v_i&=u_i+a, \quad i=1,\dots,N.
\end{aligned}
\end{equation}
For this network, the following proposition holds:
\begin{prop}
The solutions of all diffusively coupled FHN systems in the network \eqref{f7} are ultimately bounded.
\end{prop}
This fact is adopted from \cite{STE09}. The proof is based on the semi-passivity property of FHN system.

Now find the equilibrium points of this network. From the second equation of \eqref{f7} we get that $u_i=-a$, $\forall i=1,\dots,N$. Substituting $u_i$ to the first equation of \eqref{f7} we get that coupling terms dissapear and $v_i=a^3/3-a$, $\forall i=1,\dots,N$. Thus, this network has the unique equilibrium point, which is given by $x^*=\col(u_1^*,v_1^*,\dots,u_N^*,v_N^*)$, where $u_i^*=-a$, $v_i^*=a^3/3-a$, $\forall i=1,\dots,N$. Linearizing Eq.~\eqref{f7} around the equilibrium point $x^*$ we get:
\begin{equation}\label{f8}
\begin{aligned}
\epsilon \dot u_i&=(1-a^2)u_i-v_i+\sum\limits_{j=1}^{N}\gamma_{ij}\left[u_j-u_i\right], \\
\dot v_i&=u_i, \quad i=1,\dots,N.
\end{aligned}
\end{equation}

Introduce matrices $E_\epsilon$ and $A$ and vector \newline
$x=\col(v_1,\dots,v_N,u_1,\dots,u_N)$ to present the network \eqref{f8} in a matrix form:
$$
E_\epsilon=\begin{bmatrix} I_N &0\\0&\epsilon I_N\end{bmatrix},~A=\begin{bmatrix}0_N&I_N\\-I_N& (1-a^2)I_N-L(G)\end{bmatrix}.
$$
Thus, the system \eqref{f8} can be rewritten as:
$$
E_\epsilon\dot x = A x.
$$
This system is singularly perturbed. If matrix $-A$ is Hurwitz, then all the eigenvalues of matrix $A$ have positive real parts. To check this fact use the Proposition~\ref{sing}:
\begin{enumerate}
\item The ``fast'' matrix $F=L(G)-(1-a^2)I_N$ should be Hurwitz.
\item The ``slow'' matrix $S=(L(G)-(1-a^2)I_N)^{-1}$ should be Hurwitz.
\end{enumerate}
Since $S=F^{-1}$ we should only check if the matrix $S$ is Hurwitz. We consider the case of undirected graph, i.e. Laplacian matrix $L(G)$ is symmetric, therefore there exist matrix $P$ such that
$$
P^{-1}L(G)P=\Lambda=\diag(0,\lambda_2,\dots,\lambda_n),
$$
while
$$
P^{-1}SP=\Lambda-(1-a^2)I_N.
$$
This matrix is Hurwitz if $1-a^2-\lambda_{\max}(L(G))>0$. Thus, all conditions of Theorem~\ref{des} are fulfilled and the following theorem holds:
\begin{thm}\label{FHNt}
FHN network with diffusive coupling \eqref{f7} and undirected graph is desynchronized by output $u=\col(u_1,\dots,u_N)$ if $1-a^2-\lambda_{\max}(L(G))>0$ and its initial condition is not the equilibrium point $x^*=\col(u_1^*,v_1^*,\dots,u_N^*,v_N^*)$, where $u_i^*=-a$, $v_i^*=a^3/3-a$, $\forall i=1,\dots,N$.
\end{thm}
\begin{cor}
If the inequality $1-a^2-\lambda_{\max}(L(G))>0$ is fulfilled, then the  inequality $1-a^2>0$ is fulfilled too. This means that all uncoupled FHN systems in the network \eqref{f7} are oscillatory.
\end{cor}
From the obtained inequality we can conclude that the maximal eigenvalue of the Laplacian matrix should be small for the network desynchronization. This means that the overall coupling strength should be small enough and there should be few enough connections between the nodes.

\begin{figure}
\includegraphics[width=1\linewidth]{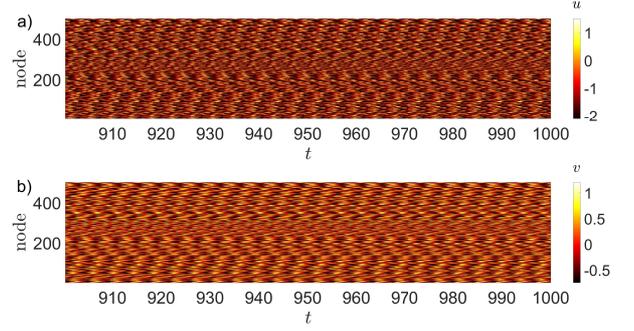}
\caption{Dynamics of diffusively coupled FHN network of $N=500$ nodes with undirected graph \eqref{f7}. (a) and (b): dynamics of the activators and inhibitors of all nodes, respectively. Parameters: $\epsilon=0.1$, $a=0.9$, $N=500$, $\lambda_{\max}(L(G))=0.00847$. Initial conditions: $u_i(0)$, $v_i(0)$, $i=1,\dots,N$ have uniform distribution on the interval $[-2;2]$.}
\label{sim}
\end{figure}

Now we consider the FHN network of $N=500$ nodes with undirected graph. Choose $\epsilon=0.1$, $a=0.9$, which means that uncoupled system is oscillatory. As  adjacency matrix consider a symmetric sparse matrix with density $0.3$, which means that it has approximately $0.3N^2$ nonzero entries. The nonzero entries are normally distributed on the interval $(0;0.005)$. The maximal eigenvalue $\lambda_{\max}(L(G))$ of the corresponding Laplacian matrix is equal to $0.0847$. Therefore the inequality $1-a^2-\lambda_{\max}(L(G))=0.1053>0$ from the Theorem~\ref{FHNt} is fulfilled.

Figure~\ref{sim} presents the simulation results. For Fig.~\ref{sim}(a),(b) $x$-axis corresponds to the time, while $y$-axis corresponds to the node number. The values of state variables of the nodes are marked by the color. The presented time interval is $[900;1000]$. One can see that both activators and inhibitors are desynchronized. Thus, the simulation results confirmed the correctness of the obtained Theorem~\ref{FHNt}.

\section{Conclusion}\label{sec:conclusion}
This paper studies the desynchronization problem in nonlinear oscillatory networks with diffusive coupling. We have introduced a new desynchronization notion for the network in terms of Yakubovich oscillatority for synchronization error system. The oscillatory network is called desynchronized if the corresponding synchronization error system is oscillatory in the sense of Yakubovich. Further, we have formulated and proved the theorem about desynchronization conditions for this network. As an example, we have considered the network of diffusively coupled FitzHugh-Nagumo systems with undirected graph. Based on the obtained theorem we have derived the simple desynchronization condition for this network, which depends on the threshold parameter and maximal eigenvalue of the Laplacian matrix. The simulation results are matched with theorem statement. The obtained results can be implemented to different types of oscillatory networks, for instance for networks of other neuron models.

\bibliography{Plotnikovbib}

\end{document}